\documentclass[twocolumn,aps,epsfig,nofootinbib]{revtex4}

\usepackage{graphicx}
\usepackage{epstopdf}
\usepackage{latexsym}
\usepackage{amssymb}
\usepackage{amsmath}
\usepackage{color}
\usepackage{mathrsfs}
\usepackage[center]{subfigure}

\begin{document}

\newcommand{\bq}{\begin{equation}}
\newcommand{\eq}{\end{equation}}
\newcommand{\bqn}{\begin{eqnarray}}
\newcommand{\eqn}{\end{eqnarray}}
\newcommand{\nb}{\nonumber}
\newcommand{\lb}{\label}
\newcommand{\PRL}{Phys. Rev. Lett.}
\newcommand{\PL}{Phys. Lett.}
\newcommand{\PR}{Phys. Rev.}
\newcommand{\PRD}{Phys. Rev. D}
\newcommand{\CQG}{Class. Quantum Grav.}
\newcommand{\JCAP}{J. Cosmol. Astropart. Phys.}
\newcommand{\JHEP}{J. High. Energy. Phys.}
\newcommand{\PLB}{Phys. Lett. B}

\title{Universal horizons and Hawking radiation in nonprojectable  2d Ho\v{r}ava gravity coupled with a non-relativistic scalar field}

\author{Bao-Fei Li$^{a, b}$}
\email{Bao-Fei_Li@baylor.edu}

\author{Madhurima Bhattacharjee${}^{a, b}$}
\email{Madhurima_Bhattacharjee@baylor.edu}
 
\author{Anzhong Wang$^{a, b}$\footnote{The corresponding author}}
\email{Anzhong_Wang@baylor.edu}

\affiliation{$^{a}$ Institute  for Advanced Physics $\&$ Mathematics, Zhejiang University of Technology, Hangzhou 310032,  China\\
$^{b}$ GCAP-CASPER, Physics Department, Baylor University, Waco, TX 76798-7316, USA}

\date{\today}

\begin{abstract}

In this paper, we study the non-projectable 2d Ho\v{r}ava gravity coupled with a non-relativistic scalar field, where the coupling is in general
non-minimal and of the form $f(\phi)R$, where $f(\phi)$ is an arbitrary function of the scalar field $\phi$, and $R$ denotes the 2d Ricci scalar.
In particular, we first investigate the Hamiltonian structure, and show that there are two-first and two-second class constraints, similar to the 
pure gravity case, but now the local degree of freedom is one, due to the presence of the scalar field. Then, we present various exact stationary 
solutions of this coupled system, and find that some of them represent black holes but now with universal horizons as their boundaries. At these 
horizons, the Hawking radiations are thermal with temperatures proportional to their surface gravities, which normally depend on the non-linear 
dispersion relations of the particles radiated, similar to the (3+1)-dimensional case.

\end{abstract}

 
\maketitle
 
\section{Introduction}
\renewcommand{\theequation}{1.\arabic{equation}} \setcounter{equation}{0}

Quantization of gravity is a subject of intense study over half a century \cite{QGs}, and various candidates have been proposed, such as
string/M-Theory \cite{strings}, Loop Quantum Gravity (LQG) \cite{LQGs},  Causal Dynamical  Triangulation (CDT) \cite{CDTs},  and Asymptotic Safety
\cite{ASs},  to name only a few of them. For more details, see  \cite{Bojowald15}. However, our understanding on each of them  is still highly limited. In 
particular, it is not clear how they are related (if there exists any),  and which is {\em the theory} we have been looking for over these years.   
One of the main reasons is the absence of experimental evidences for quantum gravitational effects. In certain senses, this is  understandable, 
considering the fact that quantum gravitational  effects are  normally expected to become important only 
at the Planck  scale, which currently  is well above the range of any man-made terrestrial experiments. However,   the situation has   been changing 
recently  with the arrival of precision cosmology \cite{KWs}.  Particularly, it was shown lately that one of the approaches adopted in loop quantum 
cosmology already leads to inconsistency with current  observations under certain circumstances  \cite{Grain16}. 

It is well known that general relativity is perturbatively not renormalizable, and its ultraviolet (UV) behavior can be dramatically changed by including high-order
derivative operators, such as the term $R_{\mu\nu}R^{\mu\nu}$ \cite{Stelle}, where $R_{\mu\nu}$ denotes the four-dimensional Ricci tensor. However, the inclusion
of such terms inevitably leads to the presence of ghosts, which makes the theory not unitary \cite{Stelle}. This problem has been plagued there since its discovery,
and has not been resolved so far. The existence of the ghosts is closely  related to the fact that the theory now contains time-derivatives with orders higher than two.  
 As a matter of fact, there exists a powerful theorem due to Mikhail Vasilevich Ostrogradsky,  who established it   in 1850  \cite{Ostrogradsky}. 
 The theorem basically states that {\em a system is not (kinematically) stable if it is described by a non-degenerate higher-order time-derivative Lagrangian}. For a recent
 introduction of this theorem, we refer readers to \cite{Woodard15}.

To avoid the Ostrogradsky ghost problem,  Ho\v{r}ava recently proposed a theory of gravity \cite{Horava}, in which Lorentz invariance (LI)
 is broken in the  UV  but recovered (approximately)  later in  the infrared (IR). Once LI is broken, one can include only high-order spatial derivative operators into the Lagrangian, 
 so the UV behavior can be dramatically improved, while  the time derivative operators  are still kept to the second-order, in order to evade Ostrogradsky's ghosts 
 and keep the theory unitary.  There are many ways to break LI. But,  Ho\v{r}ava chose to break it by considering anisotropic scaling between time and space, 
 \bq
\lb{0.1}
t \rightarrow b^{-z} t,\;\;\; x^i \rightarrow b^{-1}{x}^i,\; (i = 1, 2, ..., d)
\eq
where $z$ denotes the dynamical critical exponent, and LI requires $z = 1$, while power-counting renomalizibality requires $ z \ge d$, where $d$ denotes the 
spatial dimension of the spacetime  \cite{Horava,Visser}.   Clearly, such a scaling breaks explicitly the LI and hence $4$-dimensional diffeomorphism
invariance. Ho\v{r}ava assumed that it is broken only down to the level
\begin{equation}
\lb{0.2}
 t \to  \xi_0(t), \quad {x}^i \to  \xi^i\left(t, x^k\right),
\end{equation}
so the spatial diffeomorphism still remains. The above symmetry is often referred to as {\em the  foliation-preserving diffeomorphism},  denoted  by
Diff(M, ${\cal{F}}$).  In the original incarnation of Ho\v{r}ava gravity \cite{Horava}, the theory suffered several problems, including instability in the IR, 
strong coupling and inconsistency with observations \cite{LMP}. Since then, various modifications have been proposed, and for a recently updated review we refer readers to \cite{Wang17}. 

Among several important issues, quantization of Ho\v{r}ava gravity has been considered only in some particular cases,  despite the vast literature on  the theory.
In particular, in (3+1)-dimensional spacetimes with the projectability and detailed balance conditions, the renormalizability of Ho\v{r}ava gravity was shown to reduce to the 
one of the corresponding (2+1)-dimensional topologically massive gravity \cite{OR09}. The latter is expected to be
renormalizable   \cite{DY90}, although a rigorous  proof is still absent.   Lately, it was shown that the theory is renormalizable even without  the detailed balance condition,
by properly  choosing a gauge that ensures the correct anisotropic scaling of the propagators and their uniform falloff at large frequencies 
 and momenta \cite{BBHSS}.  
 
 Along a similar line, together with their collaborators,   two of the current authors studied the  quantization of Ho\v{r}ava gravity both with and without the projectability condition in (1+1)-dimensional (2D) spacetimes \cite{Li14,Li16}. 
Due to the   foliation-preserving diffeomorphism, the theory is non-trivial even in 2d spacetimes, in contrast to the relativistic case   \cite{Jackiw85,Brown88,GKV02},   
 although the total degree of freedom of the theory is still zero \cite{Li14,Li16}. In particular, in the projectable case, when only gravity is present, the system can be quantized by following the canonical 
 Dirac quantization \cite{Dirac64}, and the corresponding wavefunction is normalizable  \cite{Li14}.  It is remarkable to note that in this case the corresponding Hamilton can be written 
 in terms of a simple harmonic oscillator, whereby the quantization can be carried out quantum mechanically in the standard way.  When   minimally coupled to a scalar field, the momentum constraint can
 be solved explicitly in the case where the fundamental variables are functions of time only. In this case, the coupled system can also be quantized by following the Dirac process, and the corresponding 
 wavefunction is also normalizable.   
 
 In the non-projectable   case, the analysis of the 2D Hamiltonian structure shows that there are two first-class and two second-class constraints \cite{Li16}. Then, 
 following Dirac one can quantize the theory by   first requiring that the two second-class constraints be strongly equal to zero, which can be carried out by replacing the Poisson bracket by the Dirac bracket  
 \cite{Dirac64}. The two first-class constraints give rise to the  Wheeler-DeWitt equations. A remarkable feature is that orderings of the operators from a classical Hamilton to a quantum mechanical one play a 
 fundamental role in order for the Wheeler-DeWitt equation to have nontrivial solutions. In addition, the space-time is well quantized, even when it is classically singular.
 
Moreover, it was also shown  that  the 2d projectable Ho\v{r}ava gravity is exactly equal to the 2d CDT \cite{AGSW}. Such studies were further generalized to the case coupled with a scalar field \cite{AGJZ}.
In addition, the quantization of 2d Friedmann-Robertson-Walker universe was studied in \cite{VB16,Pitelli16}. 
 
In this paper, we continue our investigations in  2d   Ho\v{r}ava gravity with the non-projectable condition, but focus ourselves on two related issues: the existence of universal horizons and their Hawking radiations. 
The existence of black holes in gravitational theories with LI is closely related to the existence of light-cones \cite{HE73}. Then, in theories  in which LI is broken, it was expected that black holes should not exist,
as particles in such theories can have speeds larger than that of light, and such particles are always able to cross event horizons  to escape to infinity, even they are trapped inside them initially. 
Therefore, it was very surprised to discover that black holes exist even in such theories, but now with universal horizons as the boundaries of black holes \cite{BS11,BJS}, instead of Killing horizons  \cite{HE73}.

Since then,   universal horizons and their thermodynamics  have  been studied intensively (See, for example, \cite{Wang17} and references therein). In particular, it was showed that  universal horizons  exist  
in the three well-known black hole solutions:  the Schwarzschild, Schwarzschild anti-de Sitter, and  Reissner-Nordstr\"om \cite{LGSW}, which are also solutions of  Ho\v{r}ava gravity \cite{GLLSW}. 
At the universal horizon, the first law of black hole mechanics exists for the neutral Einstein-aether black holes  \cite{BBMa}, provided that the surface gravity is defined by \cite{CLMV}, 
\bqn
 \lb{0.3}
\kappa_{UH} \equiv  \frac{1}{2} u^{\alpha} D_{\alpha} \left(u_{\lambda} \zeta^{\lambda}\right),
 \eqn
which was obtained by considering the peering behavior of ray trajectories of constant khronon field $\phi$. However, for the charged Einstein-aether black holes,  such a first law is still absent  \cite{DWW}. 
The universal horizon radiates as a blackbody at a fixed temperature  \cite{BBMb}.   
However, different species of particles, in general, experience different temperatures \cite{DWWZ},
\bq
\lb{0.4}
T_{UH}^{z\ge 2} = \frac{2(z-1)}{z}\left(\frac{\kappa_{UH}}{2\pi}\right),
\eq
where $\kappa_{UH}$ is the surface gravity calculated from Eq.(\ref{0.3}) and $z$ is the exponent of the dominant term in the UV.  When  $z = 2$ we have
the standard result, 
$T_{UH}^{z = 2} = \frac{\kappa_{UH}}{2\pi}$,
 which was first obtained in \cite{BBMb,CLMV}.  
 
 Recently,  more careful studies of ray trajectories showed that the surface gravity for particles with a non-relativistic dispersion relation   is indeed given by \cite{DL16},
\bq
\lb{0.6}
\kappa_{UH}^{z\ge 2} = \frac{2(z-1)}{z}\kappa_{UH}.
\eq
The same results were also obtained in \cite{Cropp16}. It is remarkable to note that in terms of $\kappa_{UH}^{z\ge 2}$
and $T_{UH}^{z\ge 2}$, the standard relationship between the temperature and surface gravity of a black hole still holds here.

In this paper, we shall study  universal horizons and their thermodynamics in  2d   non-projectable Ho\v{r}ava gravity,   coupled with a 
non-relativistic scalar field. Specifically, the paper is  organized as follows: In Sec. II, we present  the general action of the coupled system and derive the corresponding 
Hamiltonian structure and field equations. In Sec. III, we find  various diagonal and non-diagonal stationary solutions of the coupled system, and correct some typos presented  
in \cite{BBC15}. In Sec. IV we first study the existence of  universal horizons in a representative spacetime found in Sec. III, and then study its
Hawking radiation by using the Hamilton-Jacobi  method. To compare it with the relativistic case, Hawking radiation at Killing horizons is also studied
in this section.  The paper is ended in Sec. V, in which we present our main conclusions.  

Before proceeding further, we would like to note that the existence of universal horizons is closely related to the existence of a globally defined time-like khronon field $\varphi$
 \cite{Wang17}. Then, all the particles are assumed to move in the increasing direction  of $\varphi $. At the beginning, universal horizons were studied in the framework of the 
 Einstein-aether theory with spherical symmetry, in which the time-like aether naturally plays the role of the  khronon field   \cite{BS11,BJS}. To generalize such conceptions to other theories,
 including  Ho\v{r}ava gravity, in which the aether field is not part of the theory,  one can consider the khronon field as a test field \cite{LACW}, a role similar to a Killing
 vector field $\xi_{\mu}$, which satisfies the Killing equations, $\nabla_{(\nu}\xi_{\mu)}  = 0$, on a given spacetime background $g_{\mu\nu}$. 
 In this paper, we shall adopt this generalization, and assume that the test khronon field satisfies the same equations as the aether field, the most general second-order partial differential equations in terms of the aether four-velocity  \cite{JM01}. For more detail, we refer readers to \cite{Wang17} and references therein.

\section{2d Ho\v{r}ava gravity coupled with a scalar field}
\renewcommand{\theequation}{2.\arabic{equation}} \setcounter{equation}{0}

The general gravitational action of Ho\v{r}ava gravity is given by,
\bq
\lb{eq1}
S_{HL}=\zeta^2 \int {dt\, dx\, N\sqrt{g} \left({\cal{L}}_{K} - {\cal{L}}_{V}\right)},
\eq
where   $\zeta^2$ denotes the coupling constant of  Ho\v{r}ava gravity, $N$   the lapse function in the Arnowitt-Deser-Misner (ADM) decomposition \cite{ADM}, and
 $g \equiv {\mbox{det}}(g_{ij})$, here $g_{ij}$ is the spatial metric defined on the leaves $t=$ Constant.   ${\cal{L}}_{K}$ is the kinetic part of the action, given by
\bq
\lb{eq2}
{\cal{L}}_{K} = K_{ij}K^{ij} - \lambda K^2,
\eq
where $\lambda$ is a dimensionless constant, and $K_{ij}$ denotes the extrinsic curvature tensor of the leaves $t=$ constant, given by 
\bq
\lb{eq3}
K_{ij}=\frac{1}{2N}\left(-\dot g_{ij}+\nabla_iN_j+\nabla_jN_i\right),
\eq
and   $K \equiv g^{ij}K_{ij}$. Here $\dot{g}_{ij} \equiv \partial{g}_{ij}/\partial t$, $\nabla_i$ denotes the covariant derivative of the metric $g_{ij}$, and $N^i$ the shift vector,
with $N_i \equiv g_{ij}N^j$.  ${\cal{L}}_{V}$ denotes the potential part of the action, and in 2d spacetimes, it takes the form \cite{Li16}, 
\bq
\lb{eq4}
{\cal{L}}_{V} = 2\Lambda - \beta a_i a^i,  
\eq
where $\Lambda$ denotes the cosmological constant, and $\beta$ is another dimensionless coupling constant. 

On the other hand, the action for a non-relativistic  scalar field takes the form,
\bqn
\lb{eq5}
S_{\phi}&=& \int {dt\, dx\, N\sqrt{g} \Bigg\{\frac{1}{2}\left(\partial_{\perp}\phi\right)^2 - \alpha_0\left(\nabla_i\phi\right)^2}\nb\\
&& ~~~~~~~~~~~~~~~~~~~~ {  - V(\phi) - f(\phi) R\Bigg\}},
\eqn
where $\partial_{\perp} \equiv N^{-1} (\partial_t - N^i\nabla_i)$,   $\alpha_0$ is a dimensionless coupling constant. In the relativistic case, it is equal to $1/2$. 
The function $f(\phi)$ is arbitrary and depends on $\phi$ only, and $R$ denotes the  Ricci scalar of the 2d spacetimes.   The total action is
\bq
\lb{eq6}
S = S_{HL} + S_{\phi} = \zeta^2 \int {dt\, dx\, N\sqrt{g} {\cal{L}}}.
\eq

\subsection{Hamiltonian Structure}
 
 The 2d spacetimes are described by the general metric,
\bq
\lb{metric}
ds^2=-N^2dt^2+\gamma^2\big(dx+N^1dt\big)^2,
\eq
subjected to the gauge freedom (\ref{0.2}), where $N, N^1$ and $ \gamma$ are in general functions of $t$ and $x$.
To be as much general as possible, we shall not impose any gauge conditions
 in this section.   Then,    the action (\ref{eq1}) takes the form,
 \bq
\lb{1.2}
S_{HL}=\int {dt dx   N \gamma \left[(1-\lambda)K^2-2{\Lambda} + \beta a_1a^1\right]}.
\eq
where  $a_1=\left(\ln N\right)'$,   and
\bq
\lb{1.3a}
K=-\frac{1}{N}\left(\frac{\dot{\gamma}}{\gamma}-\frac{N'_1}{\gamma^2}+\frac{N_1\gamma'}{\gamma^3} \right),
\eq
with  $\gamma' \equiv \partial\gamma/\partial x$, etc.  In terms of $N, N_1$ and $\gamma$, the matter action takes the form
\bqn
\lb{1.4}
S_{\phi}&=&\int dt dx N \gamma \left. \Big\{\frac{1}{2N^2}\left(\dot{\phi}-\frac{N_1 \phi'}{\gamma^2}\right)^2-\frac{\alpha_0}{\gamma^2}\phi'^2 \right.\nb\\ &&\left.
- V(\phi)- f(\phi)R \right.\Big\},
\eqn
where
\bq
\lb{1.5}
R=\frac{2}{N\gamma}\left[ \partial_\mu(N\gamma n^{\mu} K)-\left(\frac{N'}{\gamma}\right)' \right].
\eq 
Here $n^\mu \equiv N^{-1}(1, - N^1)$ denotes the normal vector to the hypersurfaces $t = $ Constant. 
Then, we find 
\bqn
\lb{1.6}
\pi_{N}&\equiv&\frac{\partial \mathcal{L}}{\partial\dot{N}} =  0,\quad
\pi_{N_1}\equiv\frac{\partial \mathcal{L}}{\partial\dot{N_1}} =  0,\nb\\
\pi&=&\frac{\partial \mathcal{L}}{\partial\dot{\gamma}}=2K(\lambda-1)-2f'\frac{\dot{\phi}}{N}+2f'\frac{\phi' N_1}{N\gamma^2},\nb\\
\pi_{\phi}&=&\frac{\partial \mathcal{L}}{\partial\dot{\phi}}=\frac{\gamma}{N}(\dot{\phi}-N_1\frac{\phi'}{\gamma^2})+2f'\gamma K.
\eqn
  After a Legendre transformation, it can be shown that  the Hamiltonian can be cast in the form, 
\bq
\lb{1.7}
\mathcal{H}_0=N\mathcal{H}+N_1\mathcal{H}^1-2\beta\left(\frac{N'}{\gamma}\right)',
\eq
where 
\bqn
\lb{1.8}
\mathcal{H}^1&=&-\frac{\pi'}{\gamma}+\frac{\pi_\phi \phi'}{\gamma^2},\\
\lb{1.81}
\mathcal{H}&=&-\frac{\pi_\phi \pi}{2 f'}+\frac{(\lambda-1)\pi_\phi}{f'}K+(1-\lambda)K^2\gamma\nb\\
&& +2\Lambda \gamma+\alpha_0\frac{\phi'^2}{\gamma} -\frac{\gamma}{2}\left(\frac{\pi_\phi}{\gamma}-2f'K\right)^2\nb\\
&&+\gamma V(\phi)-2\left(\frac{f'\phi'}{\gamma}\right)'\nb\\
&&+\beta \frac{N'^2}{N \gamma}+2\beta\left(\frac{N'}{N\gamma}\right)'.
\eqn
Here  K can be expressed in terms of the canonical fields and their momenta, 
\bq
K=\frac{\pi \gamma+2f'\pi_\phi}{4\gamma f'^2-2\gamma(1-\lambda)}.
\eq
A straightforward evaluation of poisson brackets between momentum constraints shows 
\bq
\lb{1.9}
\Big\{\mathcal{H}^1(x),\mathcal{H}^1(x')\Big\}=\left(\frac{\mathcal{H}^1(x')}{\gamma^2(x')}+\frac{\mathcal{H}^1(x)}{\gamma^2(x)}\right)\partial_{x'}\delta(x-x'), 
\eq
which is the same as in the pure gravity case \cite{Li16}. The poisson bracket between $\mathcal{H}$ and $\mathcal{H}^1$ will not vanish on the constraint surface 
because of the appearance of  terms related to the lapse function $N$ in the Hamiltonian constraint $\mathcal{H}$. Therefore, we need to redefine the momentum 
constraint by adding a term proportional to the
primary constraint $\pi_N$, which generates the diffeomorphisms of $N$,
\bq
\lb{1.10}
\tilde{\mathcal{H}}^1=\mathcal{H}^1+\frac{N'}{\gamma^2}\pi_N.
\eq
In principle, one can also add a term generating diffeomorphisms of $N_1$. However,   in the present  case, since the Hamiltonian constraint doesn't depend on $N_1$, this term is not mandatory. 
In terms of $\tilde{\mathcal{H}}^1$, the structure of Eq.(\ref{1.9}) will not change, while one can show that  $\tilde{\mathcal{H}}^1$ now commutes with $\mathcal{H}$  on the constraint surface, 
\bqn
\lb{1.11}
\Big\{\tilde{\mathcal{H}}^1(x),\mathcal{H}(x')\Big\}&=&-\left(4 \rm c \pi+\frac{2\rm{b} \pi_\phi}{\gamma}\right)\tilde{\mathcal{H}}^1(x)\delta(x-x')\nb \\&&+\frac{\mathcal{H}(x)}{\gamma^2(x)}\partial_x\delta(x-x').
\eqn
Here
 $\rm c\equiv -\alpha/2-2\xi^2\alpha^2$ and $\rm b \equiv \alpha \xi(2\beta-1)-\frac{1}{2\xi}[1+2\alpha(1-\lambda)]$, where $\alpha^{-1} \equiv 4 \xi^2+2(\lambda-1)$. Note that in writing down the 
 above expression, we had set $f(\phi)=\xi \phi$ for the sake of simplicity.  Thus,  the total Hamiltonian of the coupled system can be written as
\bq
\lb{1.12}
\mathcal{H}_{t}=N\mathcal{H}+N_1\tilde{\mathcal{H}}^1+\sigma \pi_N+\sigma_1 \pi_{N_1}.
\eq
For this coupled system, there are two first-class constraints $\tilde{\mathcal{H}}^1$ and $\pi_{N_1}$,
 and two second-class constraints $\mathcal{H}$ and $\pi_N$. 
 
Note that no other constraints will be generated by the equations of motion (E.O.M.) of the said four constraints because the secondary constraint $\tilde{\mathcal{H}}^1$ will not give rise to any tertiary constraints due to Eqs.(\ref{1.9}) and (\ref{1.11}), 
while on the other hand the preservation of $\mathcal{H}$ will only produce two differential equations for lapse function $N$ and Lagrange multiplier 
$\sigma$ since $\mathcal{H}$ is a second-class constraint. Thus, the Dirac procedure of finding all the constraints in the Hamiltonian formulation terminates 
at the level of secondary constraints,  and the physical degrees of freedom in the configuration space is one which is due to the introduction of the scalar 
 field into the whole system, while in the pure gravity case it is zero \cite{Li16}.

\subsection{Field Equations}

The variations of the total action $S$ with respect to $N, N_1, \gamma$ and $\phi$, yield, respectively,
\bqn
\lb{eq7a}
 &&(1-\lambda)\gamma K^2+2\beta\left(\frac{N'}{N\gamma}\right)'+\frac{\beta N'^2}{N^2\gamma}+\gamma\left(2\Lambda +V\right)\nb\\
 && ~~~~~ +\frac{\gamma}{2N^2}\left(\dot \phi-\frac{N_1\phi'}{\gamma^2}\right)^2+\frac{\alpha_0\phi'^2}{\gamma}\nb\\
 && ~~~~~ +\frac{2K}{N}\left(f'\dot \phi\gamma-\frac{f'\phi'N_1}{\gamma}\right)-\left(\frac{2f'\phi'}{\gamma}\right)'= 0,\nb\\
\eqn
\bqn
\lb{eq7b}
&&\frac{2(1-\lambda)K'}{\gamma}+\frac{\phi'}{N\gamma}\left(\dot \phi-\frac{N_1\phi'}{\gamma^2}\right)\nb\\
&&~~~~~ +\frac{2f'\phi'K}{\gamma} +\left(\frac{2f'\dot \phi}{N\gamma} -\frac{2f'\phi'N_1}{N\gamma^3}\right)'\nb\\
&& ~~~~~ +\frac{2\gamma'}{N\gamma^3}\left(f'\dot\phi\gamma-\frac{f'\phi'N_1}{\gamma}\right)=0,
\eqn
\bqn
\lb{eq7c}
&& 2(1-\lambda)\left(\dot K +\frac{N_1K'}{\gamma^2}-\frac{NK^2}{2}\right)-\frac{\beta N'^2}{N\gamma^2}\nb\\
&&~~~~~ +\frac{1}{2N}\left(\dot \phi-\frac{N_1\phi'}{\gamma^2}\right)^2+\frac{2N_1\phi'}{N\gamma^2}\left(\dot \phi-\frac{N_1\phi'}{\gamma^2}\right)\nb\\
&&~~~~~ -N(2\Lambda +V) +2f'\dot\phi K+2f'\phi'\frac{N_1K}{\gamma^2} \nb\\
&&~~~~~ +2f'\phi'\frac{N'}{\gamma^2} +\alpha_0\phi'^2\frac{N}{\gamma^2} -2K\left(f'\dot \phi-\frac{f'\phi'N_1}{\gamma^2}\right)\nb\\
&& ~~~~~ +\left(\frac{2f'\dot\phi\gamma}{N\gamma}-\frac{2f'\phi'N_1}{N\gamma^2}\right)_{,t} \nb\\
&& ~~~~~ -\frac{2N_1'}{\gamma^2}\left(f'\dot\phi\gamma-f'\phi'\frac{N_1}{\gamma}\right)\nb\\
&& ~~~~~ +\frac{4N_1\gamma'}{N\gamma^4}\left(f'\dot\phi\gamma-f'\phi'\frac{N_1}{\gamma}\right)\nb\\
&&~~~~~ +\left(\frac{2N_1f'\dot\phi}{N\gamma^2}-\frac{4f'\phi' N_1^2}{N\gamma^4}\right)'=0,
\eqn
\bqn
\lb{eq7d}
&&\left(\frac{\gamma\dot \phi}{N}-\frac{N_1\phi'}{N\gamma}\right)_{,t} -\left(\frac{N_1\dot \phi}{N\gamma}-\frac{N_1^2 \phi'}{N \gamma^3}\right)'-2\alpha_0\left(\frac{N\phi'}{\gamma}\right)'\nb\\
&&+N\gamma V'-2f''\dot\phi\gamma K+2(f'\gamma K)^.+2f'' \phi' \frac{N_1K}{\gamma}\nb\\&&
-2\left(\frac{f'N_1K}{\gamma}\right)'+2f''\phi'\frac{N'}{\gamma}-2\left(\frac{f'N'}{\gamma}\right)'=0.
\eqn
Here $f'(\phi) \equiv df(\phi)/d\phi$, etc.  Note Eqs.(\ref{eq7a})-(\ref{eq7d}) hold for any function $f(\phi)$. 
 
\section{Stationary Spacetimes }
\renewcommand{\theequation}{3.\arabic{equation}} \setcounter{equation}{0}

In this section, we will study stationary spacetimes of the 2d Ho\v{r}ava gravity coupled with a non-relativistic scalar field, presented in the last section. 
Setting all the time derivative terms to zero in Eqs.(\ref{eq7a})-(\ref{eq7d}), and
\bq
\lb{eq7e}
f(\phi) = \xi \phi,
\eq
where $\xi$ is a constant, we find that
\bqn
\lb{eq8a}
 &&(1-\lambda)\gamma K^2+2\beta\left(\frac{N'}{N\gamma}\right)'+\frac{\beta N'^2}{N^2\gamma}+\frac{N_1^2\phi'^2}{2N^2\gamma^3}\nb\\
 &&~~~~~~~~~~~~~~~ +\frac{\alpha_0\phi'^2}{\gamma}+\gamma(2\Lambda + V) -\frac{2K\xi\phi'N_1}{N\gamma}\nb\\
&& ~~~~~~~~~~~~~~~ -\left(\frac{2\xi\phi'}{\gamma}\right)'= 0, 
\eqn
\bqn
\lb{eq8b}
&&\frac{2(1-\lambda)K'}{\gamma}-\frac{N_1\phi'^2}{N\gamma^3}+\frac{2\xi\phi'K}{\gamma}\nb\\
&&~~~~~~~~~~~~~~~ -\left(\frac{2\xi\phi'N_1}{N\gamma^3}\right)' -\frac{2\xi\phi'\gamma'N_1}{N\gamma^4}=0,
\eqn
\bqn
\lb{eq8c}
&& 2(1-\lambda)\left(\frac{N_1K'}{\gamma^2}-\frac{NK^2}{2}\right)-\frac{\beta N'^2}{N\gamma^2}-\frac{3N_1^2\phi'^2}{2N\gamma^4}\nb\\
&&~~~~~~~~~~~~~~~ +\alpha_0\phi'^2\frac{N}{\gamma^2}+4\xi\phi'\frac{N_1K}{\gamma^2}+2\xi\phi'\frac{N'}{\gamma^2}\nb\\
&&~~~~~~~~~~~~~~~ -N(2\Lambda+V)+\frac{2\xi N_1'\phi'N_1}{\gamma^3} \nb\\
&&~~~~~~~~~~~~~~~ -\frac{4\xi\gamma'\phi'N_1^2}{N\gamma^5}-\left(\frac{4\xi\phi' N_1^2}{N\gamma^4}\right)'=0,
\eqn
\bqn
\lb{eq8d}
&&\left(\frac{N_1^2 \phi'}{N \gamma^3}\right)'-2\alpha_0\left(\frac{N\phi'}{\gamma}\right)'+N\gamma V'\nb\\
&& ~~~~~~~~~~~~~~~ -2\xi\left(\frac{N_1K}{\gamma}\right)' -2\xi\left(\frac{N'}{\gamma}\right)'=0.
\eqn

\subsection{Diagonal Solutions}

When the metric is diagonal, we have  
\bq
\lb{eq8e}
N_1 = 0, 
\eq
so the extrinsic curvature $K$ vanishes and Eq.(\ref{eq8b}) holds identically, while  Eqs.(\ref{eq8a}), (\ref{eq8c}) and (\ref{eq8d}) reduce, respectively, to
\bqn
\lb{eq9a}
&& 2\beta \left(\nu'' - \nu'\mu'\right)   + {\beta} {\nu'}^2  - 2\xi \left(\phi'' - {\phi'\mu'}\right)   + \alpha_0 {\phi'}^2 \nb\\
&& ~~~~~~~~~~~~~~~~~~ = -  (V + 2\Lambda)e^{2\mu},\\
\lb{eq9b}
&& {\beta}  {\nu'}^2    - 2\xi {\phi'\nu'}   - \alpha_0 {\phi'}^2  = - (V + 2\Lambda)e^{2\mu},\\
\lb{eq9c}
&& 2\xi\left(\nu'' + {\nu'}^2 - {\nu'\mu'}\right) + 2\alpha_0\left(\phi'' - {\phi'\mu'} + \nu'\phi'\right)   \nb\\
&& ~~~~~~~~~~~~~~ =  e^{2\mu} V', 
\eqn
where $\nu \equiv \ln N$ and $\mu \equiv \ln \gamma$. 

It should be noted that static diagonal solutions were  studied recently in \cite{BBC15} with $\Lambda = 0 = \xi$. However, comparing the above equation (\ref{eq9a}) with Eq.(12) given in \cite{BBC15}, it can be seen that
the second-order derivative term $\nu''$ (or $N''$) is missing there. This is because, when taking the variation of the total action with respect to $N$, the authors of \cite{BBC15} incorrectly assumed that $a_1$ is independent 
of $N$. Unfortunately, as a result, all the solutions resulted from Eq.(12) given in \cite{BBC15} in general are not solutions of the field equations of the 2d Ho\v{r}ava gravity coupled with a non-relativistic scalar field. 

Using the gauge freedom given by Eq.(\ref{0.2}), without loss of the generality, we can always set $\mu = -\nu$, that is,
\bq
\lb{eq10}
 N = \frac{1}{\gamma} = e^{\nu}.
 \eq
To solve Eqs.(\ref{eq9a})-(\ref{eq9c}), let us further consider the case where $V = -2\Lambda$, so that Eqs.(\ref{eq9a}) - (\ref{eq9c}) reduce to,
\bqn
\lb{eq11a}
&& 2\beta \left(\nu'' + {\nu'}^2 \right)   + {\beta} {\nu'}^2  - 2\xi \left(\phi'' + {\phi'\nu'}\right)   \nb\\
&& ~~~~~~~~~~~~~~~~~~ + \alpha_0 {\phi'}^2   = 0,\\
\lb{eq11b}
&& {\beta}  {\nu'}^2    - 2\xi {\phi'\nu'}   - \alpha_0 {\phi'}^2  = 0,\\
\lb{eq11c}
&& \nu'' + 2{\nu'}^2  + \frac{\alpha_0}{\xi}\left(\phi'' + 2\nu'\phi'\right)    =  0.
\eqn
Then, from Eqs.(\ref{eq11a}) and (\ref{eq11b}) we find that
\bq
\lb{eq12}
 \nu'' + 2{\nu'}^2  - \frac{\xi}{\beta} \left(\phi'' + 2\nu'\phi'\right)    =  0.
\eq
Thus, Eqs.(\ref{eq11c}) and (\ref{eq12}) show that there are two possibilities,
\bq
\lb{eq13}
 (i)\; {\alpha_0}\beta + {\xi}^2  \not= 0;  \quad (ii)\; {\alpha_0}\beta + {\xi}^2 =  0.
\eq
 
 \subsubsection{$ {\alpha_0}\beta + {\xi}^2  \not= 0$}
 
 In this case we must have 
 \bqn
 \lb{eq14a}
 && \nu'' + 2{\nu'}^2 = 0,\\
  \lb{eq14b}
 && \phi'' + 2{\nu'}\phi' = 0,
 \eqn
 which have the solutions,
 \bqn
 \lb{eq15}
 N &=& \sqrt{C_0 x + C_1}, \nb\\
 \phi &=&  \phi_0\ln\left(C_0 x + C_1\right) + \phi_1,
 \eqn
 where $C_i$ and $\phi_i$ are the integration constants. Without loss of the generality, we can always set $C_0 =1$, so the metric and scalar field finally take the form,
 \bqn
 \lb{eq16} 
 ds^2 &=& -\left(x - x_0\right)dt^2 + \frac{dx^2}{x - x_0},\nb\\
  \phi &=& \phi_0\ln\left(x - x_0\right) + \phi_1,
  \eqn
  where $x_0 \equiv - C_1$. Clearly, the scalar field is singular at $x = x_0$, so is the corresponding spacetime.

 \subsubsection{$ {\alpha_0}\beta + {\xi}^2  = 0$}
 
 In this case, there are only two independent equations which are Eqs.(\ref{eq11b}) and (\ref{eq11c}). Now if substituting the relation $\alpha_0=-\xi^2/\beta$
  into these equations and defining a new constant $\kappa=\xi/\beta$, one can easily arrive at,
\bqn
\lb{eq16.1}
\nu'^2-2\kappa\phi'\nu'+\kappa^2\phi'^2&=&0,\\
\lb{eq16.2}
\nu''+\nu'^2-\kappa\phi''-\kappa^2\phi'^2&=&0.
\eqn
 The first equation tells us that $\nu'$ and $\phi'$ are linearly dependent, that is,
\bq
\lb{eq16.3}
\nu=\frac{\xi}{\beta}\left( \phi - \phi_0\right),
\eq
 which also makes the second equation hold identically, where $\phi_0$ is a constant. Therefore, in the current case {\em for any chosen  $\phi$, the solution (\ref{eq16.3}) will satisfy the field
 equations (\ref{eq11a})-(\ref{eq11c})}. The corresponding metric takes the form,
 \bqn
 \lb{eq16.4}
 ds^2 &=& - e^{\frac{2\xi(\phi-\phi_0)}{\beta}}dt^2 + e^{-\frac{2\xi(\phi-\phi_0)}{\beta}}dx^2,
 \eqn
 for ${\alpha_0} = -  {\xi}^2/\beta$.

 \subsection{Non-diagonal Solutions}
 
 In this case, using the gauge transformations (\ref{0.2}), without loss of generality, we can always set
 \bq
 \lb{eq17}
 \gamma = 1,
 \eq
 so the metric takes the form,  
\bq
\lb{2.1}
ds^2=-N^2(x)dt^2+\big(dx+h(x)dt\big)^2.
\eq
Then, Eqs.(\ref{eq8a})-(\ref{eq8d}) reduce to
\bqn
\lb{eq17a}
 &&(1-\lambda)K^2+2\beta\left(\frac{N'}{N}\right)'+\frac{\beta N'^2}{N^2}+2\Lambda+V(\phi) +\frac{h^2\phi'^2}{2N^2}\nb\\&&
~~~~~~~~~+\alpha_0\phi'^2-\frac{2K\xi\phi'h}{N}-2\xi\phi''= 0,\\
\lb{eq17b}
&&2(1-\lambda)K'-\frac{h\phi'^2}{N}+2\xi\phi'K-\left(\frac{2\xi\phi'h}{N}\right)'=0, \\
\lb{eq17c}
&& 2(1-\lambda)\left(hK'-\frac{NK^2}{2}\right)-\frac{\beta N'^2}{N} -\frac{3h^2\phi'^2}{2N}\nb\\
&&~~~~~~~~~ -N(2\Lambda +V) +\alpha_0\phi'^2 N +4\xi\phi'h K\nb\\
&&~~~~~~~~~ +2\xi\phi'N'+2\xi h'\phi'h -\left(\frac{4\xi\phi' h^2}{N}\right)'=0, \\
\lb{eq17d}
&&\left(\frac{h^2 \phi'}{N}\right)'-2\alpha_0(N\phi')'+NV'-2\xi(hK)'-2\xi N''=0,\nb\\
\eqn
where
\bq
\lb{eq17e}
K = \frac{h'}{N}.
\eq
 To solve the above equations, in the following we shall consider some particular cases.  
 
 \subsubsection{$N(x) = 1$}

 In this case, let us first consider the solution with $\phi = \phi_0$, where $\phi_0$ is a constant. Then, from Eq.(\ref{eq17a}) we find that
\bq
\lb{2.2}
{h'}^2=\frac{2\hat\Lambda}{\lambda-1},
\eq
where $\hat \Lambda \equiv \Lambda + V(\phi_0)/2$. The above equation has the solution,
\bq
\lb{eq18}
h(x)=\pm \sqrt{\frac{2\hat\Lambda}{\lambda-1}} \; x=\pm \eta x. 
\eq
It can be shown that in this case a  killing horizon exists,  located at $x_{KH} = \pm \eta^{-1}$.

\subsubsection{$\xi = 0$}

When $\xi = 0$, Eqs.(\ref{eq17a})-(\ref{eq17d}) reduces to
\bqn
\lb{eq19a}
 &&(1-\lambda)\left(\frac{h'}{N}\right)^2+2\beta\left(\frac{N'}{N}\right)'+\frac{\beta N'^2}{N^2} + \hat{V} \nb\\
 &&
~~~~~~~~~ +\alpha_0\phi'^2 +\frac{h^2\phi'^2}{2N^2} = 0,\\
\lb{eq19b}
&&2(1-\lambda)\left(\frac{h''}{N} - \frac{h'N'}{N^2}\right) -\frac{h\phi'^2}{N} =0, \\ 
\lb{eq19c}
&& 2(1-\lambda)\left(\frac{h''}{N} - \frac{h'N'}{N^2} - \frac{{h'}^2}{2hN}\right)-\frac{\beta N'^2}{hN} -\frac{3h\phi'^2}{2N}\nb\\
&&~~~~~~~~~ +\frac{N}{h}\left(\alpha_0\phi'^2 - \hat{V}\right)  =0, \\
\lb{eq19d}
&&\left(\frac{h^2 \phi'}{N}\right)'-2\alpha_0(N\phi')'+N\hat{V}' =0,
\eqn
where $\hat{V} \equiv V + 2\Lambda$. To solve the above equations, let us consider the case,   
\bq
\lb{eq20}
N = h, \quad \hat{V} = 0,
\eq
for which the above equations reduce to
\bqn
\lb{eq21a}
 &&2\beta \nu'' + (1-\lambda +\beta){\nu'}^2 = - \frac{1+2\alpha_0}{2} {\phi'}^2,\\
\lb{eq21b}
&&2(1-\lambda)\nu'' = {\phi'}^2, \\ 
\lb{eq21c}
&& 2(1-\lambda)\nu''  -  (1-\lambda +\beta){\nu'}^2 = \frac{3-2\alpha_0}{2}   {\phi'}^2, ~~~~\\
\lb{eq21d}
&&(1-2\alpha_0)\left(e^{\nu}\phi'\right)' = 0,
\eqn
where $\nu = \ln N$.
  To solve the above equations, let us consider the cases 
$\alpha_0 = 1/2$ and $\alpha_0 \not= 1/2$, separately.   

{\bf Case B.2.1) $\alpha_0 = 1/2$:} This is  the relativistic case, and Eq.(\ref{eq21d}) is satisfied identically, while from Eqs.(\ref{eq21a}) and (\ref{eq21c}), we find 
\bq
(1-\lambda+\beta)\nu''=0.
\eq
If   $\lambda\not=\beta +1$, it can be shown that the above equations have only the trivial solution  in which $\nu$ and $\phi$ are all constants. 
On the other hand,  when $\lambda=\beta +1$, Eqs.(\ref{eq21a})-(\ref{eq21c}) reduce
to a single equation,  
\bq
\lb{eq22}
2\beta \nu'' = -\phi'^2, \; (\beta = \lambda -1).
\eq
for the two arbitrary functions $\nu$ and $\phi$.  Again, similar to Case A.2 considered in the last subsection, the solutions are not uniquely determined. In fact, for any given $\phi$, the solution,
\bq
\lb{eq23}
\nu(x) = -\frac{1}{2\beta}\int^{x}{dx'\int^{x'}{{\phi'}^2(x'')dx''}} + C_1x + C_0,
\eq
 will satisfy the field equations (\ref{eq21a}) and (\ref{eq21c}), where $C_1$ and $C_0$ are two integration constants.

 {\bf Case B.2.2) $\alpha_0 \not= 1/2$:} In this case, from Eq.(\ref{eq21d}) we find
\bq
\lb{eq24}
\phi'=C_0 e^{-\nu},
\eq
where $C_0$ is another  constant. Substituting it into Eq.(\ref{eq21b}), we obtain
\bq
\lb{eq25}
NN''-N'^2+{\cal{D}}=0, 
\eq
where ${\cal{D}} \equiv {C_0^2}/({2(\lambda-1)})$.  The above equation has two particular solutions,
\bqn
N_A(x) &=&\frac{1}{2C_1^2}e^{C_1(x+C_2)}-\frac{{\cal{D}} }{2}e^{-C_1(x+C_2)},\\
N_B(x)&=&\frac{1}{2C_1^2}e^{-C_1(x+C_2)}-\frac{{\cal{D}} }{2}e^{C_1(x+C_2)},
\eqn
where $C_1$ and $C_2$ are two integration constants. Correspondingly, the scalar field $\phi$ is given, respectively, by,
\bqn
\phi_A(x)&=&-\frac{2}{\sqrt{{\cal{D}} }} \tanh^{-1} \left(\frac{e^{C_1(C_2+x)}}{\sqrt{{\cal{D}} }C_1}\right),\\
\phi_B(x)&=&\frac{2}{\sqrt{{\cal{D}}}} \tanh^{-1}\left(C_1\sqrt{\cal{D}}  e^{C_1(C_2+x)}\right).
\eqn

\section{Universal Horizons and Hawking Radiation}
\renewcommand{\theequation}{4.\arabic{equation}} \setcounter{equation}{0}

In this section, we shall consider two issues, universal horizons and the corresponding  Hawking radiations. 
As a representative case, we shall 
focus on the solution given by Eqs.(\ref{2.1}) and (\ref{eq18}) with $N = 1$. Without loss of the generality, we consider only the case with ``-" sign, that is, 
\bqn
\lb{2.3aa}
ds^2 &=& - dt^2 + \left(dx - \eta x dt\right)^2\nb\\
&=& - \left(1-\eta^2x^2\right)dt^2 - 2\eta x dt dx + dx^2,
\eqn
where $ - \infty < t, x < \infty$. 
The corresponding inverse metric is given by
\bq
\lb{imetric}
g^{tt} = -1, \;\;\; g^{tx} = - \eta x, \;\;\; g^{xx} = 1 - \eta^2 x^2,
\eq
which is non-singular, except at the infinities $x = \pm \infty$. The latter are coordinate singularities, similar to the 4d de Sitter space. In fact, 
the extrinsic curvature and 2d Ricci scalar are  all finite, and given by $-\eta$ and $2\eta^2$, respectively. However, there exist two cosmological Killing horizons, located, 
respectively,  at $x_{KH} = \pm\eta^{-1}$.  Similar to the 4d de Sitter space, the time-translation Killing vector, $\xi^{\mu} = \delta^{\mu}_{t}$,  is time-like only in the 
region $x^2 < x_{KH}^2$. In the regions $x^2 > x_{KH}^2$, the Killing vector becomes spacelike, and only in these regions can the universal horizon exist, as the latter is
defined by \cite{Wang17},
\bq
\lb{UH}
(\xi \cdot u) = 0. 
\eq
Since the four-velocity $u$ of the khronon field is always time-like,  Eq.(\ref{UH}) has solutions only when $\xi$
becomes spacelike, which are the regions in which  $x^2 > x_{KH}^2$ holds. 

To see the difference between the physics at Killing horizons and that at universal horizons, let us first consider Hawking radiation at the Killing horizon.

\subsection{Hawking radiation at the Killing horizon}

As shown in \cite{DWWZ}, at a Killing horizon only relativistic particles are radiated quantum mechanically. So, in this subsection we consider only 
 the relativistic limit in which the dispersion relation of radiated  massless scalar particles satisfies $k^2\equiv k_{\lambda}k^{\lambda} = 0$.
Considering only the positive outgoing particles, $k_t=-\omega < 0$, we find
\bq
\lb{2.4}
k_x^{\pm}=\frac{\omega(h\pm1)}{1-h^2}\;,
\eq 
which is singular for $k_x^+$ at the Killing horizon at which we have $h\left(x_{KH}\right) = 1$. Then, from the following formula \cite{DWWZ},
\bq
\lb{2.5}
2 \rm{Im} S=\rm{Im}\oint k^+_x dx  
=\frac{\omega}{T_{KH}},
\eq
we find that
\bq
\lb{2.5a}
T_{KH}=-\frac{h'(x_{KH})}{2\pi} =\frac{\eta}{2\pi},  
\eq
where $x_{KH} = -\eta^{-1}$.    On the other hand, the surface gravity at the Killing horizon is given by \cite{HE73},
\bqn
\lb{2.5b}
\kappa_{KH} &\equiv& \sqrt{-\frac{1}{2}\left(D_{\mu}\chi_{\nu}\right)\left(D^{\mu}\chi^{\nu}\right)}\nb\\
&=& \eta,
\eqn
where $D_{\mu}$ denotes the covariant derivative with respect to the 2d metric $g_{\mu\nu}$, and $\chi^{\mu} = \delta^{\mu}_{t}$ is the timelike Killing vector. Therefore, the standard form, 
\bq
\lb{2.5c}
T_{KH} = \frac{\kappa_{KH}}{2\pi},
\eq
holds.

\subsection{Universal Horizons and Hawking Radiation}

The existence of a universal horizon is closely related to the existence of a globally defined timelike scalar field $\varphi$ \cite{LACW,Wang17}, 
\bq
\lb{2.6}
u_\mu=\frac{\partial_\mu\varphi}{\sqrt{-g^{\alpha \beta}\partial_\alpha \varphi\partial_\beta \varphi}},\quad u_{\lambda}u^{\lambda} = -1,
\eq
where the equation of $\varphi$ is given by the action \cite{EJ},  
\bqn
\lb{2.7}
S_{u}&=& \int {dt dx N \gamma\Big[\frac{\kappa_1}{2}F^{\alpha\beta}F_{\alpha\beta}+\kappa_2(D_\alpha u^\alpha)^2}\nb\\
&& ~~~~~~~~ ~~~~~~~ {+\sigma(u^\alpha u_\alpha+1)\Big]},
\eqn
where $F_{\alpha\beta}\equiv D_\alpha u_\beta-D_\beta u_\alpha$,  $\sigma$ is a Lagrange multiplier, and $\kappa_{1,2}$ are two coupling constants.
It should be noted that the action (\ref{2.7})  remains unchanged under the transformations, 
\bq
\lb{phis}
{\varphi} = {\cal{F}}(\tilde\varphi),
\eq
where ${\cal{F}}(\tilde\varphi)$ is a monotonically increasing or decreasing function of $\tilde\varphi$ only. In the following, we shall use this property to choose
${\cal{F}}(\tilde\varphi)$ so that $d\varphi$ is along the same direction as $dt$ in the regions we are interested in.

Under the background  (\ref{2.3aa}), 
 we find that the equations of motion are given  by,  
\bqn
\lb{2.8}
\kappa_1(1-\eta^2 x^2)u_0''-\sigma u_0&=&0,\\
\lb{2.8b}
\kappa_1\eta x u_0''+\kappa_2 (u^1)''-\sigma u_1&=&0,\\
\lb{2.8c}
u_0^2+2\eta x u_0u_1-(1-\eta^2 x^2)u_1^2-1&=&0.
\eqn
Generally, these coupled non-linear equations are difficult to solve. One simple solution can be obtained   when $\kappa_1 = 0$, in which we find $\sigma u_0=0$. Since $u_0 \not= 0$ 
we must have $\sigma=0$, and Eqs.(\ref{2.8})-(\ref{2.8c}) have the solution \footnote{Eq.(\ref{2.8c}) is a quadratic equation for $u_0$, so in general it has two solutions. In the following 
we shall consider only the one with  the minus sign, as the one with the plus sign will give the same results.}, 
\bqn
\lb{2.9}
u^0&=&\frac{\eta x u^1- \sqrt{G(x)}}{\eta^2x^2-1},\;\;\;
u^1=c x+d,\nb\\
G(x) &\equiv& \left(c^2 - \eta^2\right)x^2 +2cdx + \left(d^2+1\right), 
\eqn
or inversely 
\bqn
\lb{2.10}
u_0&=&-\sqrt{G(x)},\nb\\
u_1&=&\frac{-(c x+d)+\eta x \sqrt{G(x)}}{\eta^2 x^2-1},
\eqn
where $c$ and $d$ are two integration constants. In asymptotically flat spacetimes, these two constants can be determined by requiring that \cite{BS11,LACW}: (a) it be aligned asymptotically with the time translation
Killing vector; and (b) the khronon have a regular future sound horizon. However, the spacetime we are studying is asymptotically de Sitter, and these conditions cannot be applied to the present case. Instead, we
shall leave  this possibility open, as long as it allows a globally defined khronon field $\varphi$. Since only the latter is essential for the existence of the universal horizon, as explained previously at the end of 
Introduction.
 Then, one may  ask what is their physical meanings. 
To see these, let us first calculate the quantity,
\bq
\nabla_\alpha u_\beta=c s_\alpha s_\beta+\hat{c} u_\alpha s_\beta,
\eq
where  
\bq
\hat{c}\equiv \frac{x\eta^2-c(cx+d)}{\sqrt{1+(cx+d)^2-x^2\eta^2}}.
\eq
Thus,  $c$ is directly related to the expansion of the aether. In fact, we have $\theta \equiv  g^{\alpha\beta} \nabla_{\alpha}u_{\beta} = c$.
On the other hand, assuming that the aether is moving alone the trajectory $x^{\mu} = x^{\mu}(\tau)$, where $\tau$ is the proper time measured by aether,
from Eq.(\ref{2.9}) we  find
\bq
\lb{eqd}
\left. u^1 \equiv \frac{dx(\tau)}{d\tau}\right|_{c=0} = d,
\eq
that is, the parameter $d$ is directly related to the constant part of the velocity of the aether.
 
In order to have the solution (\ref{2.9}) well-defined for all the values of $x \in (-\infty , \infty)$, we must assume that  $G(x) \ge 0$, which yields
\bq
\lb{2.11}
c^2 \ge \left(1+d^2\right)\eta^2.  
\eq
On the other hand, the   universal horizon is located at \cite{Wang17}, $\left(u \cdot \xi\right) = - \sqrt{G(x)} =  0$.  Since $G(x) \ge 0$ for $x \in (-\infty, \infty)$, we must have \cite{LGSW},
\bq
\lb{2.14a}
  G\left(x_{UH}\right) =  0, \;\;\;\; \left. \frac{dG(x)}{dx}\right|_{x = x_{UH}} = 0,
\eq
at the universal horizon $x = x_{UH}$. Inserting Eq.(\ref{2.9}) into the above equations, we find that 
\bq
\lb{solution}
c = \epsilon_c \eta\sqrt{1+d^2},\;\;\;  x_{UH} = -  \epsilon_c \frac{\sqrt{1+d^2}}{\eta d},
\eq
where $\epsilon_c = {\mbox{Sign}}(c)$. It is interesting to note that the above solution for $c$ saturates the bound of Eq.(\ref{2.11}). We also note that 
\bq
\lb{locations}
x^2_{UH} - x^2_{KH} = \frac{1}{(\eta d)^2} > 0,
\eq
as expected.  

\begin{figure}[tbp]
\centering
{\includegraphics[width=8cm]{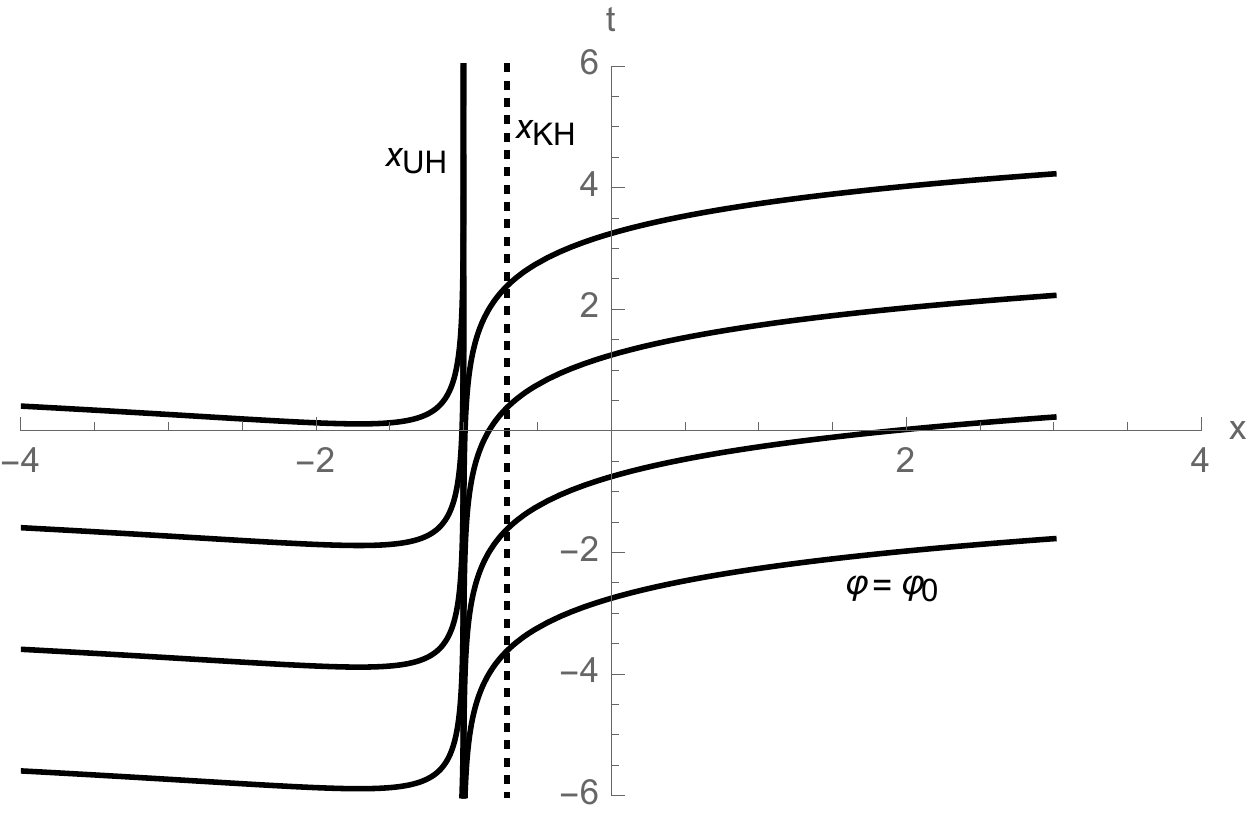}}
\caption{The curves of $\varphi = $ Constant. In this figure, we choose $\epsilon_c = 1,\; d=1$, $\eta=\sqrt{2}$.  
The universal horizon (dotted vertical line) is located at $x_{UH}=-1$, and the black vertical line denotes the location of the cosmological Killing horizon located
at $x_{KH}=-\frac{1}{\sqrt{2}}$.} 
\label{khronon}
\end{figure}

On the other hand, from Eqs.(\ref{2.6}) and (\ref{phis}), we find that the khronon field takes the form,
\bq
\lb{phiA}
\varphi=t+f(x),
\eq
where we had chosen ${\cal{F}} = - \tilde{\varphi}$, and  dropped the tilde from $\tilde{\varphi}$ for the sake of simplicity,  without causing any confusions.
The function   $f$  satisfies the differential equation,
\bq
f'(x)=\frac{u^1-\eta x \sqrt{G(x)}}{(\eta^2 x^2-1) \sqrt{G(x)}}.
\eq
In Fig. \ref{khronon}, we show the curves of  Constant $\varphi$, from  which  it can be seen
clearly the peeling behavior of  the curves of constant $\varphi$  at the universal horizon, while these curves are well-behaved across
the Killing horizon.  

From Eq.(\ref{2.9}), we can construct a spacelike unit  vector $s_{\mu} = s_0\delta^{t}_{\mu} + s_1\delta^{x}_{\mu}$, 
which is orthogonal to $u^{\mu}$. It can be shown that $s_{\mu}$ has the non-vanishing components,  
\bqn
\lb{2.12}
s_0&=&-(cx+d),\nb\\
s_1&=&\frac{\eta x u^1- \sqrt{G(x)}}{\eta^2 x^2-1}.
\eqn
Then, we can project   $k^{\mu}$ onto $u^\alpha$ and $s^\alpha$, and obtain,  
\bqn
\lb{2.13}
k_u\equiv (k\cdot u) =-\omega u^0+ k_xu^1,\nb\\
k_s\equiv (k\cdot s) =-\omega u_1-k_xu_0.
\eqn
To proceed further, we need to consider the aether four-velocity $u_{\mu}$ in the regions $x > x_{UH}$ and  $x < x_{UH}$, separately. 
In particular,  $\epsilon_c $  is set to unity in Eq.(\ref{solution}) which leads to the solution,
\bqn
\lb{2.14}
u_0&=&-|d\eta x+\sqrt{d^2+1}|,\nb\\
u_1&=&d,\nb\\
u^0&=&\sqrt{d^2+1},\nb\\
u^1&=&\eta \sqrt{d^2+1}x+d,\nb\\
f'&=&-\frac{d}{x\eta d+\sqrt{d^2+1}},\nb\\
f&=&- \frac{1}{\eta}\ln \left(\eta x d+\sqrt{d^2+1}\right),
\eqn
for  $x\textgreater x_{UH}$.  
When $x\textless x_{UH}$, we find that 
 $u_0$ and $u^1$ remain the same while $u^0$, $u_1$, $f'$ and $f$ are changed to 
\bqn
u^0&=&\frac{\eta^2x^2\sqrt{d^2+1}+2d\eta x+\sqrt{d^2+1}}{\eta^2 x^2-1},\nb\\
u_1&=&-\frac{2\eta x \sqrt{d^2+1}+d\eta^2x^2+d}{\eta^2 x^2-1},\nb\\
f'&=&\frac{d}{x\eta d+\sqrt{d^2+1}}+\frac{2\eta x}{1-x^2 \eta^2},\nb\\
f&=&\frac{1}{\eta}\ln \left(\frac{d x \eta+\sqrt{d^2+1}}{1-x^2\eta^2}\right),\; (x \textless x_{UH}).
\eqn

At the universal horizon, similar to the (3+1)-dimensional case \cite{DWWZ},  relativistic particles cannot be emitted in the form of Hawking radiation.   
 Thus, in the following we consider only the particles with the following non-relativistic 
dispersion relation  \cite{DWWZ},  
\bq
\lb{2.15}
k_u^2=k_s^2+a_2\frac{k_s^4}{k_0^2},
\eq
where $a_2$ is a dimensionless constant of order one, and $k_0$ is the cutoff energy scale. For $k \ll k_0$, the particles become relativistic. 
Then, from Eq.(\ref{2.13})  we find  
\bqn
k_u&=&-\frac{1}{u_0}(k_su^1-\omega),\nb\\
k_x&=&-\frac{1}{u_0}(\omega u_1+k_s).
\eqn
Combined with the dispersion relation (\ref{2.15}), we find that $k_s$ has a simple pole at the universal horizon $x=x_{UH}$  with $u_0(x_{UH})=0$. Thus,  we assume that  near the universal horizon 
we have
\bq
k_s=- \frac{b(x) }{u_0},
\eq
where $b(x=x_{UH})\neq 0$. To calculate the temperature given by Eq.(\ref{2.5}) but now at the universal horizon, in principle we only need  the
Laurent expansion of $k_x$ in the neighborhood  of the universal horizon. Setting $\epsilon=x-x_{UH}$, for the special case given by Eq.(\ref{2.14}),  we find
\bqn
\lb{taylor}
u_0&=&-d \eta \epsilon,\nb\\
u^1&=&-\frac{1}{d}+\epsilon \eta \sqrt{d^2+1}, \nb\\
b(x)&=&b_0+b_1\epsilon + {\cal{O}}\left(\epsilon^2\right),\nb\\
k_x&=&\frac{b_0}{\eta^2d^2\epsilon^2}+\frac{1}{\epsilon}\left(\frac{\rm \omega}{\eta}+\frac{b_1}{\eta^2 d^2}\right)+ {\cal{O}}\left(1\right),
\eqn
 for $x>x_{UH}$,  where
\bqn
b_0&=&\pm \frac{k_0}{\sqrt{a_2}d},\nb\\
b_1&=&\eta d^2 \rm \omega -\eta d b_0 \sqrt{d^2+1}.
\eqn
When $x<x_{UH}$, the Taylor expansions of $u^1$ and $b(x)$ remain the same as in Eq.(\ref{taylor}) while $u_0$ and $k_x$ are changed to
\bqn
\lb{taylor1}
u_0&=&d \eta \epsilon,\nb\\
k_x&=&\frac{b_0}{\eta^2d^2\epsilon^2}+\frac{1}{\epsilon}\left(-\frac{\rm \omega}{\eta}+\frac{b_1}{\eta^2 d^2}\right)+ {\cal{O}}\left(1\right),
\eqn
Correspondingly, with the help of dispersion relation Eq.(\ref{2.15}), one can show
\bqn
b_0&=&\pm \frac{k_0}{\sqrt{a_2}d},\nb\\
b_1&=&-\eta d^2 \rm \omega -\eta d b_0 \sqrt{d^2+1}.
\eqn
In order to figure out the temperature at the universal horizon, one needs to analytically continue the radial momentum $k_x$ to the complex plane, combining 
Eqs.(\ref{taylor}) and (\ref{taylor1}), it's easy to conclude that, by setting $x=x_{UH}+\epsilon e^{i \theta}$, for $\theta \in (0,2\pi)$
\bq
k_x=\frac{b_0}{\eta^2d^2\epsilon^2  e^{2i \theta}}+\frac{2 \omega}{\eta \epsilon  e^{i \theta}}-\frac{b_0\sqrt{d^2+1}}{\eta d \epsilon},
\eq
Then, using Eq.(\ref{2.5}),
\bq
\frac{\omega}{T_{KH}}=\rm{Im}\oint k^+_x dx=\frac{4\pi\omega}{\eta},
\eq
from which we find that, 
\bqn
T_{UH}=\frac{\eta}{4\pi}.
\eqn

The surface gravity at the  universal horizon is given by \cite{Wang17} \footnote{It should be noted that $\kappa_{UH}$ given by Eq.(\ref{SG}) can also be
obtained by considering the peeling behavior of the khronon field $\varphi$ given by Eq.(\ref{phiA}), as it was done in \cite{CLMV}.},
\bq
\lb{SG}
\kappa_{UH}=\frac{1}{2}D_u(u\cdot\zeta) = \frac{\eta}{2},
\eq
from which we find that the standard relation 
\bq
T_{UH}=\frac{\kappa_{UH}}{2\pi},
\eq
is satisfied at the universal horizon. This is similar to the (3+1)-dimensional case \cite{  CLMV,BBMb,DWWZ}. For more general case with the dispersion relation, 
\bq
\lb{2.15a}
k_u^2=k_s^2\sum_{n=0}^{2z}{a_{n}\left(\frac{k_s}{k_0}\right)^{n}},
\eq
it can be shown that the (3+1)-dimensional results \cite{DWWZ},
\bq
\lb{2.15ab}
T^{z\ge 2}_{UH}=\frac{\kappa^{z\ge 2}_{UH}}{2\pi} = \left(\frac{2(z-1)}{z}\right)\left(\frac{\kappa_{UH}}{2\pi}\right),
\eq
can be also obtained.


\section{Conclusions}

In this paper, we  studied the non-projectable Ho\v{r}ava gravity coupled with a non-relativistic scalar field, in which the coupling is in general  non-minimal through 
the interaction term $f(\phi)R$. The Hamiltonian structure of this coupled system is very similar to that of  pure gravity case. There exist two first-class constraints and two 
second-class constraints (The combinations of two second-class constraints will generate two global first-class constraints which account for global time reparametrization 
symmetry of Ho\v{r}ava gravity as first pointed out in \cite{DJ}). Therefore, the local degrees of freedom is one due to the presence of the scalar field. 

We also found diagonal static solutions for the couplings $f(\phi)=\xi \phi$, and showed that Killing horizons exist in such solutions, but  the scalar field turns out to be singular 
at these Killing horizons. For the non-diagonal stationary solutions,  when the lapse function and the spatial metric component $g_{11}$ are set to one, we found that the 
solutions represent black holes, in which both Killing and universal horizons exist. At the Killing horizon, the temperature of Hawking radiation is proportional to its surface 
gravity defined as in the relativistic case [cf. Eq.(\ref{2.5b})] \cite{HE73}. 

To study  locations of the universal horizons, we first considered a test  timelike scalar field in such a fixed background \cite{LACW}, and found solutions of  the test field, whereby
the universal horizons located at $\chi \cdot u = 0$ were found. By using the Hamilton-Jacobi  method \cite{DWWZ}, we calculated  the temperature at the universal horizon,
and found that it is proportional to the  modified surface gravity defined by Eq.(\ref{SG}).  For $z = 2$ of the dispersion relation (\ref{2.15a}), the  modified surface gravity
given by Eq.(\ref{SG}) satisfies the standard relation with its temperature, $T_{UH}={\kappa_{UH}}/({2\pi})$, similar to the (3+1)-dimensional case \cite{BBMb,CLMV}.
But, in more general cases, both of them will depend on $z$, as shown by Eq.(\ref{2.15ab}), although the standard relation,   
$T^{z\ge 2}_{UH}={\kappa^{z\ge 2}_{UH}}/({2\pi})$,  is still expected to hold \cite{DL16,Cropp16}. 

The results presented  in this paper show clearly that the existence of  universal horizons and their thermodynamics are independent of dimensions of spacetimes concerned.
Therefore, the 2d Ho\v{r}ava gravity provides an ideal place to address these important issues, which  often technically become very complicated in higher dimensional  
spacetimes.

\section*{Acknowledgements}

  Part of the work was done when  B.-F.L. and M.B. were visiting Zhejiang University of Technology (ZUT), China, and part of it was done when A.W.
 was visiting  the State University of Rio de Janeiro (UERJ), Brazil. They would like to thank ZUT and UERJ for their hospitality.  This work was supported in part by Ci\^encia Sem Fronteiras, 
 No. 004/2013 - DRI/CAPES, Brazil (A.W.); and National Natural Science Foundation of China (NNSFC), Grant Nos. 11375153 (A.W.) and No. 11675145 (A.W.). 
 B.-F.L. and M.B.  are supported by Baylor University through the physics graduate program, and partially by the NNSFC Grant Nos. 11375153 and No. 11675145. 

\end{document}